\documentclass[9pt,a4paper]{article}

\usepackage{authblk} 

\newcommand{\fcal}{ \mathcal{F} }
\newcommand{\rvec}{ \mathbf{r} }
\newcommand{\Rvec}{ \mathbf{R} }
\newcommand{\fvec}{ \mathbf{f} }

\newcommand{\rhat}{\hat{\textbf{r}}}

\usepackage[utf8]{inputenc}
\usepackage[english]{babel}
\usepackage{amsmath}
\usepackage{amsfonts}
\usepackage{amssymb}
\usepackage{graphicx}
\usepackage{fullpage}
\usepackage{comment}
\usepackage{hyperref}
\usepackage{indentfirst} 
\usepackage{authblk} 
\usepackage{comment}
\usepackage{amsfonts}
\usepackage{xcolor}
\usepackage[super,comma,sort&compress]{natbib} 

\usepackage{scalerel}
\DeclareUnicodeCharacter{2010}{-}
\DeclareUnicodeCharacter{2212}{-}
\DeclareUnicodeCharacter{2032}{ }

\author[1]{Javier Diaz}
\author[1]{Marco Pinna\thanks{mpinna@lincoln.ac.uk}}
\author[1]{Andrei V. Zvelindovsky}
\author[2,3,4]{Ignacio Pagonabarraga\thanks{ipagonabarraga@ub.edu}}
\affil[1]{Centre for Computational Physics, University of Lincoln. Brayford Pool, Lincoln, LN6 7TS, UK}
\affil[2]{Departament de Física de la Matèria Condensada, Universitat de Barcelona, Martí i Franquès 1, 08028 Barcelona, Spain
}
\affil[3]{
{CECAM, Centre Europ\'een de Calcul Atomique et Mol\'eculaire, \'Ecole Polytechnique F\'ed\'erale de Lausanne,
Batochime - Avenue Forel 2, 1015 Lausanne, Switzerland }
}
\affil[4]{Universitat de Barcelona Institute of Complex Systems (UBICS), Universitat de Barcelona, 08028 Barcelona, Spain
}

\title{
Non-spherical nanoparticles in block copolymer composites: 
nanosquares, nanorods and diamonds
}

\begin{document}
\maketitle

\begin{center}
\includegraphics[height=3.5cm]{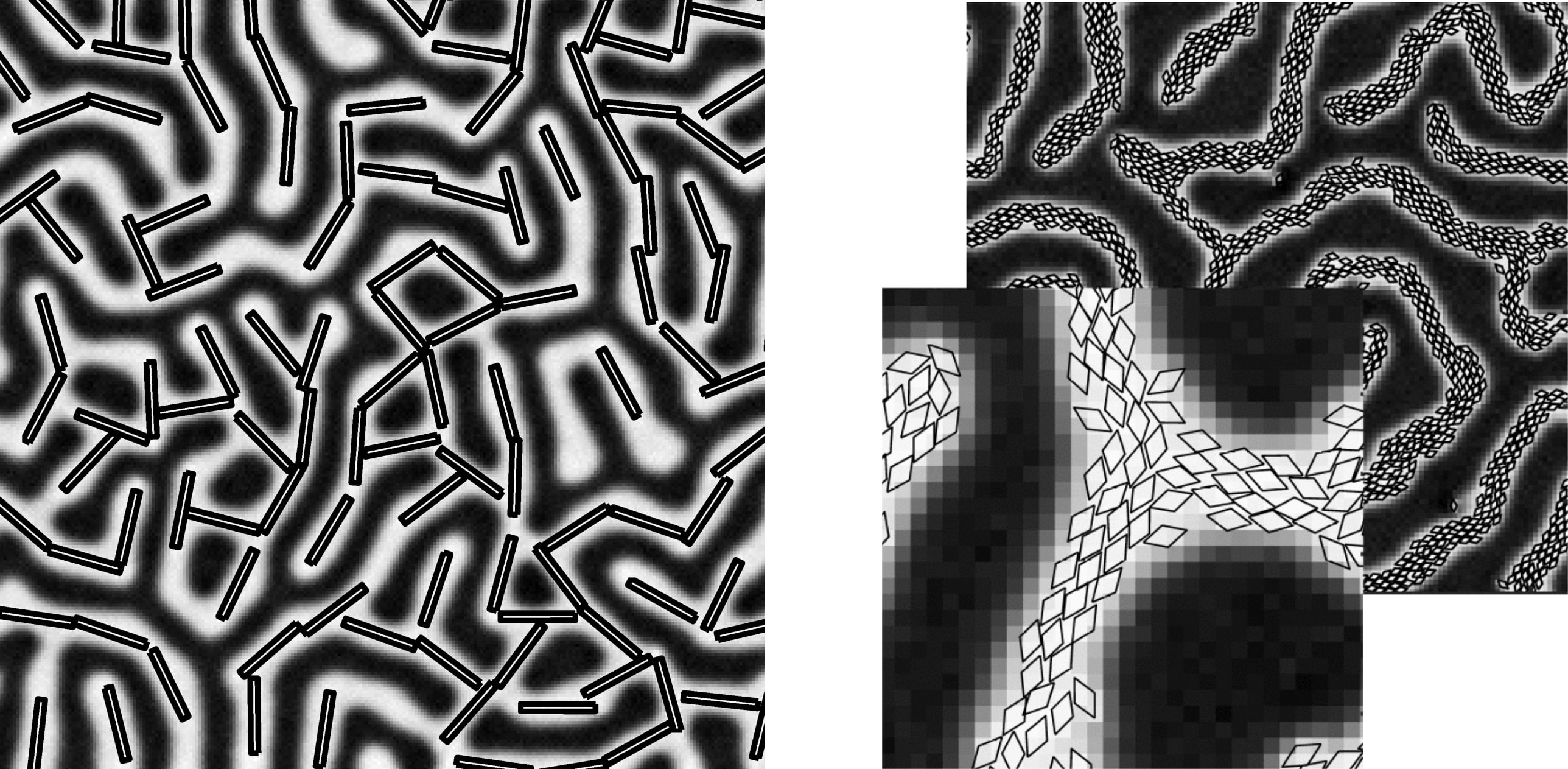}
\end{center}
\begin{center}
For Table of Contents use only
\end{center}

\begin{center}
\section*{Abstract}
\end{center}
\textit{
A hybrid block copolymer(BCP) nanocomposite computational model is proposed to study nanoparticles(NPs) with a generalised shape including squares, rectangles and rhombus. 
Simulations are used to study the role of anisotropy in the assembly of colloids within BCPs, ranging from NPs that are compatible with one phase, to neutral NPs. 
The ordering of square-like NPs into grid configurations within a minority BCP domain was investigated, as well as the alignment of nanorods in a lamellar-forming BCP, comparing the simulation results with experiments of mixtures of nanoplates and PS-\textit{b}-PMMA BCP. 
The assembly of rectangular NPs at the interface between domains resulted in alignment along the interface. 
The aspect ratio is found to play a key role on the aggregation of colloids at the interface, which leads to a distinct  co-assembly behaviour for low and high aspect ratio NPs.
}

\section{Introduction}

Polymer nanocomposite materials containing anisotropic nanoparticles have attracted a lot of attention due to their ability to create functional materials with enhanced properties\cite{hore_functional_2014}. 
Block copolymers can serve as perfect matrices to produce highly ordered polymer nanocomposite materials due to the well-defined, periodic structures that block copolymer melts have.
This periodicity makes block copolymers excellent matrices to host nanoparticles (NPs), which can be localised in specific regions of the phase-separated block copolymer\cite{okumura_nanohybrids_2000,kim_effect_2006}.

The co-assembly of block copolymer melts and colloidal nanoparticles  has long been studied with 
theoretical \cite{pryamitsyn_strong_2006,pryamitsyn_origins_2006}
and computational  
\cite{
huh_thermodynamic_2000,
thompson_block_2002,
thompson_predicting_2001,
pinna_modeling_2011,
langner_mesoscale_2012,
horechyy_nanoparticle_2014} methods, 
as well as experiments
\cite{
bockstaller_size-selective_2003,
bockstaller_block_2005,
chiu_control_2005,
kim_nanoparticle-induced_2005}. 
 The resulting nanocomposite material is not merely the addition of a passive filler. 
Instead, the properties of the polymeric matrix can be modified by the presence of colloids and the assembly of colloids is deeply intertwine with the diblock copolymer matrix\cite{ploshnik_hierarchical_2013}.
  For instance, colloids which are wetted to be compatible with one of the phases will drive the BCP into a morphological transition\cite{huh_thermodynamic_2000,lo_effect_2007,halevi_co-assembly_2014}.

 While spherically shaped colloids have been widely studied -both experimentally and theoretically \cite{langner_mesoscale_2012}- the additional difficulty of modelling anisotropy have resulted in comparably less work devoted to it. 
 Nevertheless, the orientational degree of freedom of non-spherical colloids introduces new possibilities of BCP/NP co-assembly,  given the intrinsic ordered structures of the neat block copolymer (lamellar, cylindrical, etc). 
Complex non-spherical nanoparticles such as nanorods(NR) have attracted a lot of attention as  constituents of functional polymer nanocomposite materials\cite{hore_functional_2014}.   
 For instance, gold nanorods orient along the lamellar domain when confined in one of the symmetrical phases\cite{deshmukh_two-dimensional_2007,tang_self-assembly_2009}. 
 Similarly, gold nanorods template the direction of the cylindrical domains in an asymmetrical diblock copolymer mixture\cite{laicer_gold_2005}. 
 Ordered arrays of aligned nanorods were achieved experimentally \cite{thorkelsson_direct_2012,thorkelsson_end--end_2013} in the co-assembly of BCP and highly anisotropic NPs, where NRs were organised in an end-to-end configuration. 


Theoretical and computational works have studied the self-assembly of BCP and anisotropic NP. 
Dissipative Particle Dynamics (DPD) has been largely used, thanks to the ability to combine several beads into rod-like sequences.
The phase behavior of such systems has been studied along with the orientation  of nanoparticles  \cite{he_effect_2009,he_phase_2009,he_mono-_2010} and  the effect of shear in the global orientation \cite{pan_dynamic_2011}.
 Osipov et al \cite{osipov_spatial_2016,osipov_induced_2017,osipov_phase_2018}
used Strong and Weak Segregation Limit Theory to determine the distribution of anisotropic particles in a diblock copolymer, with a low fraction of nanoparticles present in the system.  
A mesoscopic Ginzburg-Landau approach has been used to study Janus nanorods mixed with BCP\cite{guo_phase_2017}.

In this work our previous hybrid BCP/NP model \cite{pinna_modeling_2011} originally developed for spherical nanoparticles is extended  for a vast family of NP shapes.  
The shape of the NPs is motivated by experiments. 
For example, considerable attention has been given to nanocubes\cite{henzie_self-assembly_2012,wang_simple-cubic_2017} as an example of faceted particles that can lead to the formation of crystals\cite{sharma_disorder_2018}.
In this work we make use a considerably fast, coarse grained Cell Dynamic Simulation (CDS) method to describe the block copolymer dynamics, while colloidal NPs are described using  Brownian Dynamics.

The Cell Dynamic Simulation method has been used extensively both in pure block copolymer systems \cite{ren_cell_2001,pinna_large_2012,pinna_diblock_2011,dessi_cell_2013} and nanocomposite systems\cite{pinna_modeling_2011}, reproducing experiments such as aggregation of incompatible colloids \cite{diaz_cell_2017,ploshnik_hierarchical_2013}
and NP-induced phase transitions
\cite{diaz_phase_2018}. Its relative computational speed makes it suitable to study properties which involve large systems over extended times, while the phenomenological approach in its model limits its validity in the microscopic realm. This hybrid method permits to explore the high NP concentration regime, in which the presence of NPs introduces huge differences over the neat BCP matrix, such as, morphological phase transitions. 

In the following section a generalised model for anisotropic NPs will be presented. 
To our knowledge, this work represents the first mesoscopic simulations of BCP/NP assembly for cuboid, rectangular or diamond shaped nanoparticles. 
Contrary to previous works \cite{guo_phase_2017}, we aim to capture the finite width of rectangular NPs, thus allowing to study the role of the aspect ratio in the high NP concentration regime.


\section{Model}

Here we present a coarse grained model of the dynamics of BCP and NPs with a range of shapes and chemical properties. 
This model is intrinsically mesoscopic, disregarding microscopic details, i.e. the individual monomers of the polymer chains are not resolved. 
This coarse grained approach permits to achieve relatively large box sizes compared to previous works. 


The evolution of the BCP/colloids  system is determined by the excess free energy which can be separated as 
\begin{equation}
\fcal _{tot} = \fcal_{pol}+\fcal_{cc} +\fcal_{cpl}
\end{equation}
with $\fcal_{pol}$ being the free energy functional of the BCP melt, $\fcal_{cc}$ the colloid-colloid interaction and the last contribution being the coupling term between the block copolymer and the colloids.

The diblock copolymer is characterized by the order parameter $\psi ( \rvec ,t   )$ which represents the differences in the local volume fraction for the copolymer A and B 
\begin{equation}
\psi (\rvec,t )= 
\phi_A (\rvec,t)
-
\phi_B (\rvec,t)
+(1-2f_0)
\end{equation}
with respect to the relative volume fraction of A monomers in the diblock, $f_0= N_A/ (N_A +N_B)$.

The order parameter must follow the continuity equation in order to satisfy the mass conservation of the polymer: 
\begin{equation}
\frac{\partial\psi ( \rvec, t )}{\partial t}=
-\nabla\cdot \mathbf{j} (\rvec ,t ) 
\end{equation} 

If the polymer relaxes diffusely towards equilibrium, the order parameter flux can be expressed in the form 
\begin{equation}
\mathbf{j } (\rvec,t    )=
-M \ \nabla \mu (\rvec , t )
\end{equation}
as a linear function of the order parameter chemical potential
\begin{equation}
\mu (\rvec , t )=
\frac{\delta \fcal_{tot}  [ \psi] }{ \delta \psi}
\end{equation}

Introducing these equations into the continuity equation and taking into account the thermal fluctuations we obtain the Cahn-Hilliard-Cook equation (CHC)
\begin{equation}
\frac{\partial\psi ( \rvec, t )}{\partial t}=
M\ \nabla^2 \left[
\frac{\delta \fcal_{tot}  [ \psi] }{ \delta \psi}\right]
+
\xi ( \rvec, t)
\label{eq:anis.cahn}
\end{equation}
where $M$ is a phenomenological  mobility constant and $\xi$ is a white Gaussian random noise which satisfies the fluctuation-dissipation theorem\cite{ball_spinodal_1990}. 

The copolymer free energy is a functional of the local order parameter which can be expressed in terms of the thermal energy $k_B T$ as
\begin{equation}
\fcal_{pol}  [ \psi (\rvec ) ]=
\int d\rvec \left[
H(\psi) +\frac{1}{2} D | \nabla\psi  |^2   
\right]
+ \\
\frac{1}{2} B \int d\rvec  \int d\rvec' \ 
G(\rvec -\rvec ' )\psi(\rvec)\psi(\rvec') 
\end{equation}
where the first and second terms are the short and the long-range interaction terms respectively.
The coefficient $D$ is a positive constant that accounts for the cost of local polymer concentration inhomogeneities, the Green function $G(\rvec-\rvec' )$ for the laplace Equation satisfies $\nabla^2 G(\rvec-\rvec') = -\delta (\rvec-\rvec')$, $B$  is a parameter that introduces a chain-length dependence to the free energy\cite{hamley_cell_2000} and $H (\psi)$ is the local free energy \cite{hamley_cell_2000,ren_cell_2001}, 
\begin{equation}
 H(\psi )  = 
 \frac{1}{2}\left[   
 -\tau_0+ A(1-2f_0)^2
 \right]   \psi ^2  \\
 +\frac{1}{3} v (1-2f_0)\psi^3 
 +\frac{1}{4}u \psi^4
 \label{eq:anis.Hpsi}
\end{equation}
where $\tau_0,A,v,u $ are phenomenological parameters\cite{ren_cell_2001} which can be related to the block-copolymer molecular specificity. Previous works\cite{pinna_modeling_2011,ren_cell_2001,ohta_equilibrium_1986} describe the connection of these effective parameters to the BCP molecular composition.  $\tau ' = -\tau_0+A(1-2f_0)^2$, $D$ and $B$ can be expressed\cite{ohta_equilibrium_1986} in terms of degree of polymerization $N$, the segment length $b$  and the Flory-Huggins parameter $\chi$(inversely proportional to temperature)  .
 Subsequently, we will consider $u$ and $v$  constants\cite{leibler_theory_1980}, which define all the parameters identifying the BCP local  free energy $H(\psi)$ . As  previously shown \cite{sevink_selective_2011,pinna_mechanisms_2009}, CDS can be used along with more detailed approaches like dynamics self-consistent field theory (DSCFT), using CDS as a precursor in exploring parameter space due to the computationally inexpensiveness nature of CDS. We can express the time evolution of $\psi$ , Equation \ref{eq:anis.cahn}, using CDS as 
\begin{equation}  
\begin{split}
\psi ( \rvec_i , t+1   )= \psi (\rvec_i,t )-  
\delta t [ 
\langle \langle \Gamma (\rvec_i, t )\rangle\rangle 
- \Gamma (\rvec_i, t ) + \\  
B     [ 1- P (\rvec_i, t) \psi (\rvec_i,t )]   -\eta \xi (\rvec_i,  t)       ] 
]
\end{split}
\label{eq:anis.time_evol}
\end{equation}
$\rvec_i$ being the position of the node $i$ at a time $t\delta t$, and the isotropic discrete  laplacian for a quantity $X$ is given by \cite{oono_study_1988}
$\frac{1}{\delta x ^2}  [ \langle\langle X \rangle\rangle -X  ] $
where $\delta x $ is the lattice spacing. 
 Specifically, we  will use 
\begin{equation}
\langle \langle \psi \rangle \rangle = \frac{1}{6}  \sum_{NN}  \psi   +\frac{1}{12} \sum _{NNN} \psi
\end{equation} 
NN, NNN meaning nearest neighborsand next-nearest neighbors, respectively, for the two dimensional case.

In Equation \ref{eq:anis.time_evol} we have introduced the auxiliary function 
\begin{equation}
\Gamma (\rvec, t ) =
g( \psi (\rvec, t)  )- \psi (\rvec, t)+
D \left[ 
\langle\langle   \psi (\rvec, t)    \rangle \rangle   -\psi (\rvec, t)
\right]
\end{equation}
and also, the map function \cite{bahiana_cell_1990,ren_cell_2001}
\begin{equation}
g (\psi)= -\tau ' \psi -v (1-2f_0)\psi^2 -u \psi^3
\end{equation}


\subsection{Coupling between the block copolymer and nanoparticles}

In previous works\cite{pinna_modeling_2011}, the presence of $N_p$ nanoparticles has been introduced via an extra term in the free energy, given by 
\begin{equation}
F_{cpl}=
\sum_{p=1}^{N_p}
\sigma \int d\rvec\
 \psi_c(\textbf{s}) 
\left[ 
\psi(\rvec)-\psi_0
\right]^2
\label{eq:anis.fcpl.superellipse}
\end{equation}
In the case of a simple circular particle, $\mathbf{s}=
|\mathbf{s}|/R=
(x/R)^2+(y/R)^2
$
and the tagged function is related to the core/shell nature of the particle. It is a smoothly monotonically decreasing function that takes values $\psi_c(0)=1$ and $\psi_c(1)=0$. In our choice, 
\begin{equation}
\psi_c(s)=\exp \left[1-\frac{1}{1-s} \right] 
\label{eq:anis.psic}
\end{equation}
for $s<1$ and $\psi_c(s>1)=0$. This expression allows us to design a nanoparticle that interacts softly with the surrounding block copolymer, and it avoids the need to explicitly consider the NP-BCP boundary.  

\begin{figure*}[hbtp]
\centering
\includegraphics[width=1.0\linewidth]{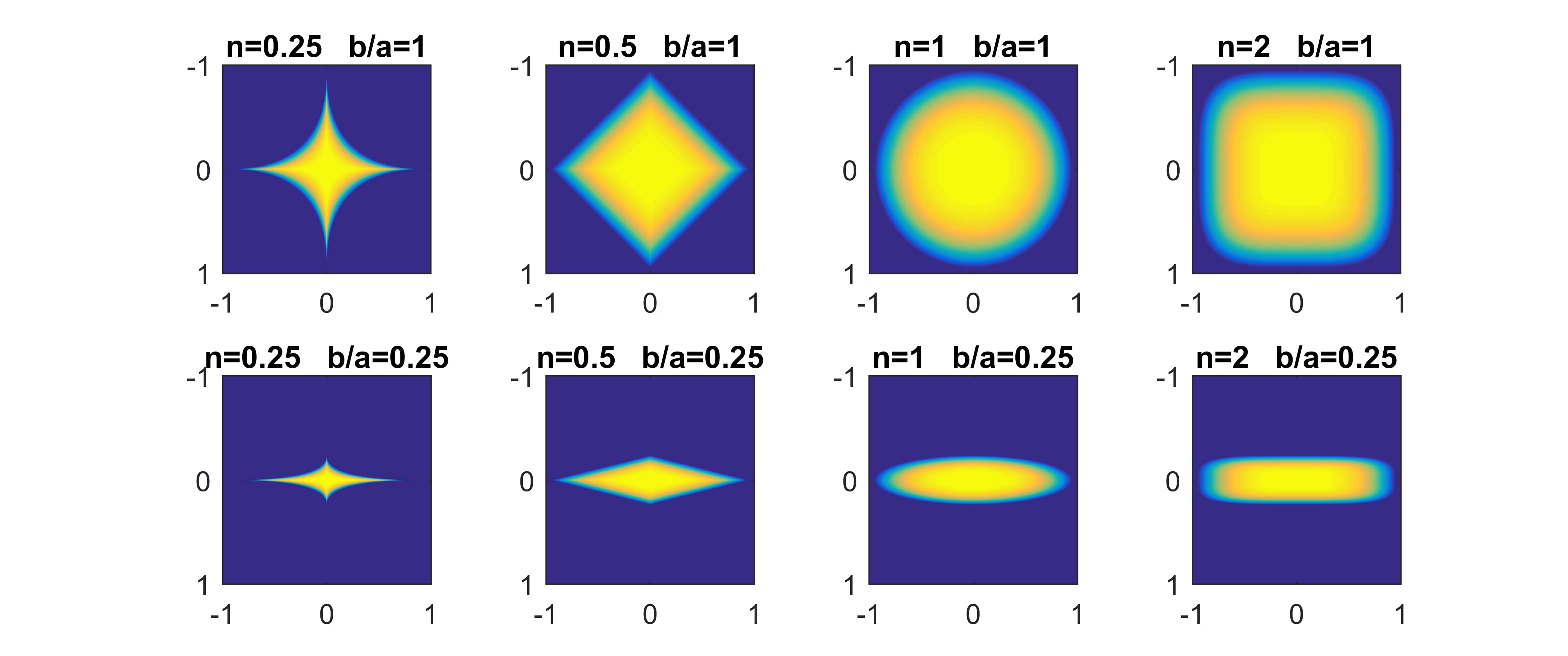}
\caption{
Colourmap of $\psi_c$ as described in Eqs. \ref{eq:anis.psic} and \ref{eq:anis.sr} for several values of $n$ and $b/a$. 
The particles are assumed to have no rotation $\phi=0$ and larger semiaxis $a=1$. 
}
\label{fig:anis.shapes}
\end{figure*}

To extend the colloidal shape from a circular particle to an elongated ellipse, one can trivially rescale the axis, such that $
s=
(x/a)^2+(y/b)^2
$
 for an ellipse that is resting horizontally (vertically) if $a>b$ ($a<b$). 
Furthermore, we can consider a generalisation of this procedure, to represent a larger family of curves, namely, superellipses. Such closed curves are given by 
\begin{equation}
s(\rvec)=\left[ \left\lvert\frac{x}{a}\right\rvert^{2n}+
 \left\lvert\frac{y}{b}\right\rvert^{2n}\right]^{1/n}  = 1 
 \label{eq:anis.sr}
\end{equation}
where the exponent $1/n$ rescales the decay\cite{donev_tetratic_2006} of $\psi_c$. 
More generally, each rotated anisotropic colloid is characterised by an unit vector $\hat{\textbf{n}}_i = (\cos\phi_i  ,\sin\phi_i )$ depending on an angle $\phi_i$. 
Again, $x,y$ should be related to the rotated $x'(\phi),y'(\phi)$ versions of the variables.  
In Fig. \ref{fig:anis.shapes} representative examples of super-ellipses are given for several values of $n$, for two values of the aspect ratio. 

The simplicity of the coupling free energy allows us to simulate a vast range of NP shapes. 
The family of superellipses in two dimensions include ellipses, rectangles, rhombus and star-like particles. 
Many of these shapes can serve as a faithful representation of experimentally-motivated non-spherical colloids such as nanocubes, nanospheres or nanorods.

\subsection{Interparticle potential}

The calculation of forces and torques requires a choice of a suitable intercolloidal pairwise additive potential. 
In the particular case of ellipsoids ($n=1$), a modified Lennard-Jones potential has been widely used\cite{gay_modification_1981}. 
Nonetheless, in this work we will study the particular case of rhomboid and rectangular shaped particles. 
To this end, we design a simplified repulsive potential that will assure non-overlapping between particles. 

We introduce a colloid-colloid contribution to the free energy as 
\begin{equation}
\fcal_{cc} = 
\sum_{i\neq j} U(\rvec_{i},\rvec_j,\phi_i,\phi_j)
\label{eq:anis.Fcc}
\end{equation}
with $U(\rvec_{i},\rvec_j,\phi_i,\phi_j)$ a potential that depends on the distance between two colloidal particles, as well as their orientations $\phi_i,\phi_j$. 
Specifically, we choose a potential that is proportional to the overlapping area (volume in three dimensions) between two colloidal particles of arbitrary shape
\begin{equation}
U(\rvec_{i},\rvec_j,\phi_i,\phi_j) =
A_\text{overlap} (\rvec_i,\rvec_j,\phi_i,\phi_j)
\label{eq:anis.Aoverlapping}
\end{equation}
which prevents overlapping between colloids.  
To our knowledge, there is no general analytic expression of the overlapping area between two parallelogram of arbitrary orientation and distance. 
Nonetheless, the Separating Axis Theorem (SAT) is widely used in computer science to test collisions between convex polygons\cite{gottschalk_obbtree:_1996}. 

In order to derive forces and torques between colloids, we require an expression of the overlapping area. 
To this end we numerically calculate the overlapping area between colloids and perform a fit of the  results into simple analytical expressions.
Sines and cosines are used to capture the appropriate  symmetry of each object. 
Additionally, the overlapping area is approximated to be proportional to the center-to-center distance.  
The occurrence of overlapping can be determined exactly and -since forces and torques are only calculated when overlapping is true- making this method computationally fast. 
The details for different shapes, orientations and distances can be found in the Supplementary Information.

\subsection{Brownian Dynamics}

Colloids undergo diffusive dynamics described by the Langevin equation in the overdamped regime. The centre of mass of each colloid follows Brownian Dynamics, that is, 
\begin{equation}
\frac{d \Rvec_i}{dt}=
\frac{1}{\gamma_t} \left( 
\fvec^{cc}_i+
\fvec^{cpl}_i+
\sqrt{2k_BT\gamma_t} \xi
\right)
\end{equation}
where $\fvec^{cpl}_i = -\partial \fcal_{cpl}/\partial \Rvec_i$ is the coupling force derived from Eq. \ref{eq:anis.fcpl.superellipse} and $\fvec^{cc}_i = -\partial \fcal_{cc}/\partial \Rvec_i$ is the colloid-colloid force derived from Eq. \ref{eq:anis.Fcc} through the fitted expression of the overlapping area in Eq. \ref{eq:anis.Aoverlapping}. 
Similarly, the orientation of particle $i$ follows 
\begin{equation}
\frac{d \phi_i}{dt} = 
\frac{1}{\gamma_R} 
\left( 
M^{cc}_i+
M^{cpl}_i+
\sqrt{2k_BT\gamma_R} \xi
\right)
\end{equation}
where $\gamma_t$ and $\gamma_R$ are the translational and rotational friction coefficients and torques can be calculated as 
$M^{cpl}_i = -\partial \fcal_{cpl}/\partial \phi_i$ and 
$M^{cc}_i = -\partial \fcal_{cc}/\partial \phi_i$, respectively for the coupling and colloid-colloid torques. 
Generally speaking, the translational diffusivity is a tensor that accounts for the anisotropic diffusivity of arbitrary shaped particles \cite{han_brownian_2006} along the parallel and perpendicular main axis of the colloid. 
Nonetheless, we are interested on the equilibrium properties of the assembly of polyhedra shaped colloids. 
Thus, we can assume these $\gamma_t$ to be a scalar\cite{tang_self-assembly_2009}.

\section{Results}


The standard values of CDS\cite{ren_cell_2001,pinna_large_2012,pinna_modeling_2011} will be used 
$
\tau_0=0.35, u=0.5,v=1.5,A=1.5,D=2.0
$
while the BCP/NP interaction is set to $\sigma=1.0$. 
The BCP time scale is set by the mobility $M=1$. 
A moderate BCP noise strength $0.1$ is chosen while the NP temperature scale is set to $k_BT=0.1$.  
A cell spacing $\delta x =1.0$ and time discretisation $\delta t=0.1$ are chosen. 
Lengths will be presented in units of grid points. 
Initially, both the position and the orientation of colloids is random. 
Furthermore, the block copolymer is initially disordered, corresponding to a quick quench at $t=0$. 

\subsection{Selective NPs}

In order to study a minimal example of anisotropic colloids in BCP, we can firstly consider square-like NPs which are compatible with the minority phase of the BCP ($\psi_0=-1$). 
By doing so, we are introducing a high effective volume fraction of colloids. 
We select a cylinder-forming BCP with $f_0=0.35$ and explore a concentration of colloids $\phi_p$ with a side length $2a_0$. 
At low concentrations, the NPs are confined within  circular BCP domains. 
At higher concentration, the NPs are effectively increasing the fraction minority monomers, enough to induce a cylinder-to-lamellae transition in the BCP. 
\cite{diaz_cell_2017,diaz_phase_2018,halevi_co-assembly_2014,kim_nanoparticle-induced_2005,lo_effect_2007,sides_hybrid_2006,huh_thermodynamic_2000}
A combination of the entropic NP-NP interaction and the enthalpic effect of the surrounding block copolymer can lead to organisation of the square-shaped colloids into a defined side-to-side assembly. 
In order to quantify an eventual grid-like arrangement, we can define the interparticle orientational parameter 
\begin{equation}
T \equiv 
\langle\langle
\left[ 2 ( \mathbf{n}_1\cdot \rhat  )^2-1 \right] ^2
\left[ 2 ( \mathbf{n}_2\cdot \rhat  )^2-1 \right] ^2  
\rangle\rangle
\end{equation}
where the brackets indicate average over all particles' neighbours. 
This parameter is $1$ for a grid-like arrangement of particles and it is clearly defined positive. 
It provides insight both on the orientational and translational order of the colloidal set of particles, as it captures the $\pi/2$ rotation symmetry of squares.

\begin{figure*}[hbtp]
\centering
\includegraphics[width=0.7\textwidth]{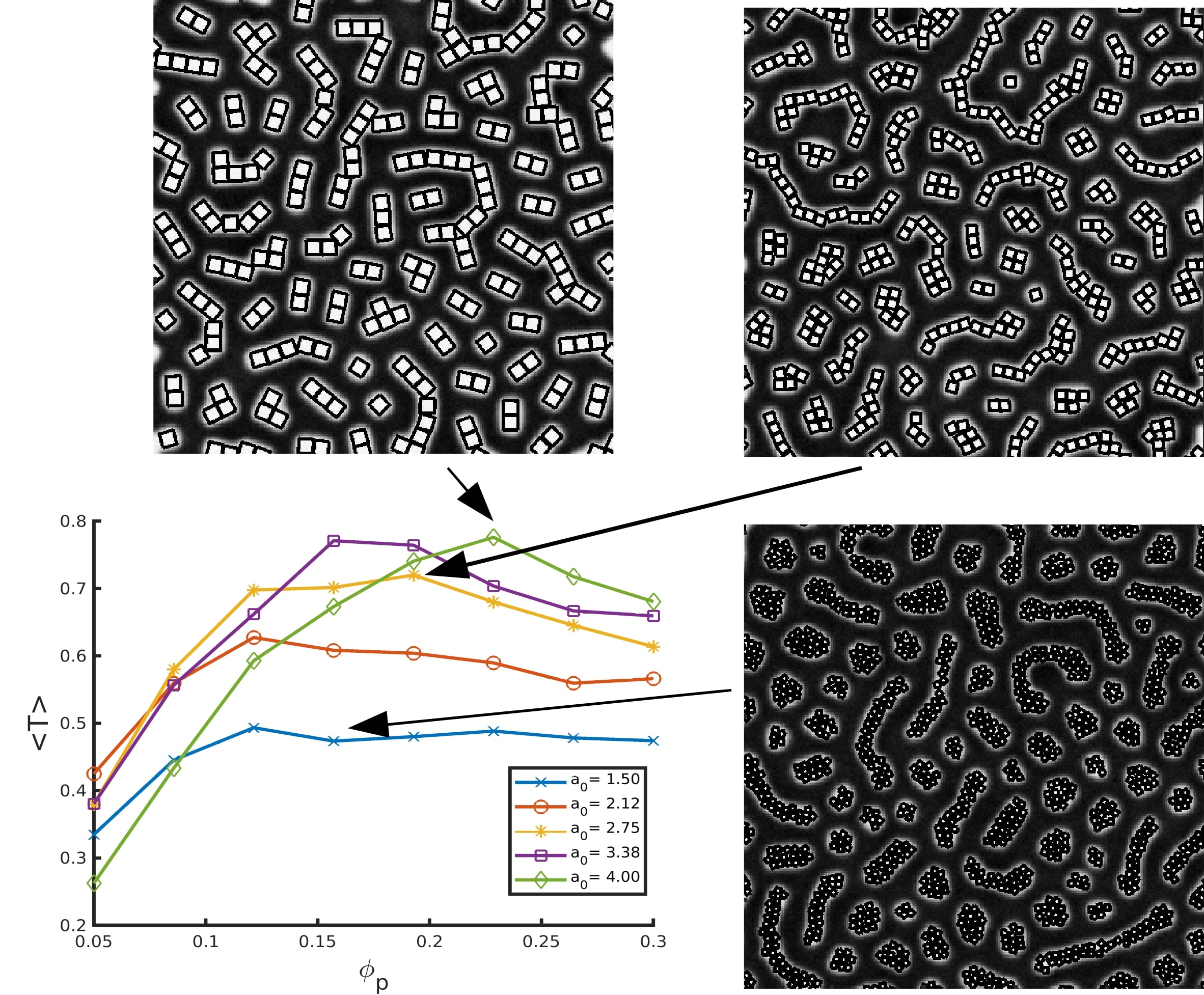}
\caption{
Interparticle orientation parameter for a concentration $\phi_p$ of square-shaped NPs for different sizes (squares with side $2a_0$).
Snapshots show representative instances of the colloidal assembly. }
\label{fig:anis.T22.squares}
\end{figure*}

We can explore the parameter space given by the fraction of particles in the system $\phi_p$ and the size of the particle $a_0$. 
By doing so we can explore  the arrangement of particles for a given level of constrain for different sizes. 
In Figure \ref{fig:anis.T22.squares} we plot such curves. 
We can observe that small-sized particles are not well oriented within the block copolymer, with $T<0.5$ for any fraction of particles when $a_0=1.5$. 
We can conclude that small particles tend to be dominated by thermal motion and neither the block copolymer, nor the entropy associate with the anisotropy of particles can induce an ordered phase. 
Larger particles, instead, can achieve ordered configurations with array-like organisation of squares. 
For a three-dimensional system, we can expect a monolayer of well-ordered cubes in a lamellar phase. 

\begin{figure*}[hbtp]
\centering
\includegraphics[width=0.9\textwidth]{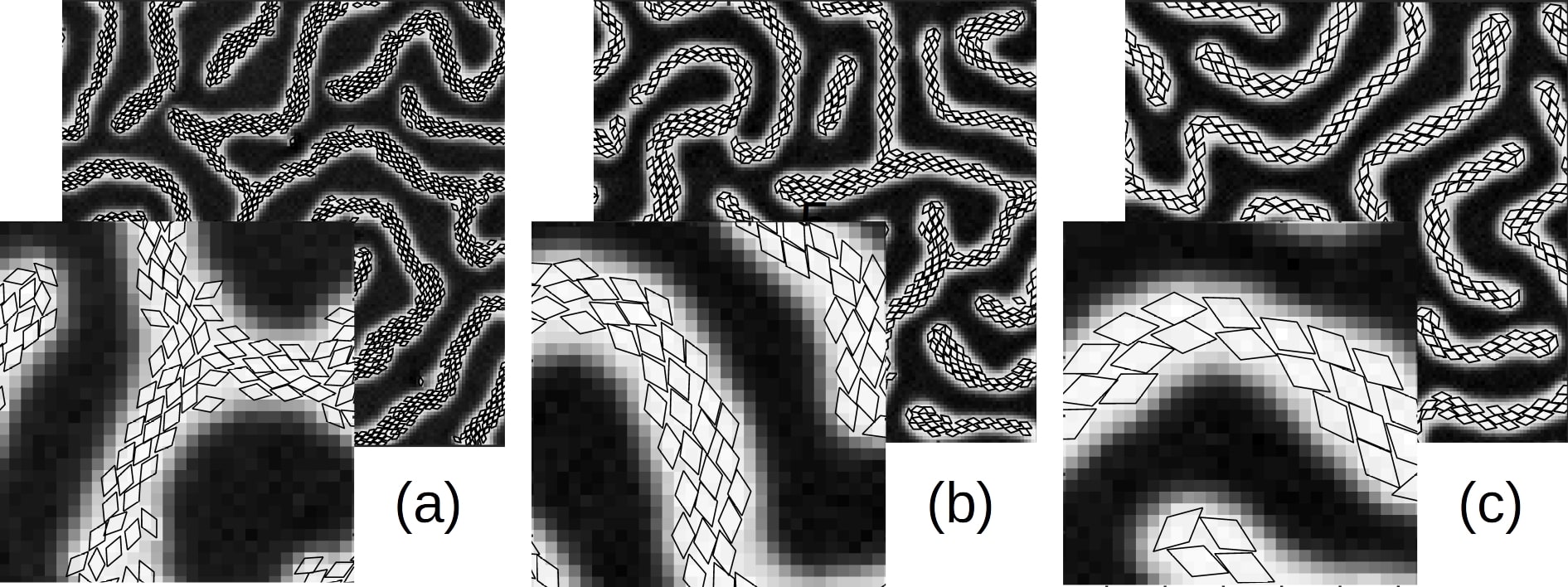}
\caption{
Rhomboid-shaped NPs compatible with the minority domain in a $f_0=0.35$ asymmetric BCP. A fixed concentration $\phi_p=0.2$ is considered for NP sizes  $a_0=\sqrt{2}, 2$ and $ 2\sqrt{2}$ for (a), (b) and (c), respectively. 
}
\label{fig:anis.diamong-sizes}
\end{figure*}

In Fig. \ref{fig:anis.diamong-sizes} we study the assembly of rhombus-shaped NPs by selecting an exponent $n=0.6$ from Eq. \ref{eq:anis.sr}. 
A value $f_0=0.35$ and $\phi_p=0.2$ can be selected, such that the BCP morphology is lamellar in the presence of a concentration of NPs. 
By doing so, rhomboid-shaped NPs segregate within the minority domains and experience a considerable local effective concentration, which leads to close-packed configuration for three different rhomboid sizes: $a_0=\sqrt{2}, 2$ and $ 2\sqrt{2}$ for (a), (b) and (c), respectively. 
The NP aspect ratio is kept constant $b/a=1/2$. 
Regardless of the size of the particles, the BCP lamellar morphology and its confinement effect leads to a highly ordered NP configuration. 

\subsection{Neutral NPs}

\begin{figure*}[h!]
 \centering
 \includegraphics[width=0.9\textwidth]{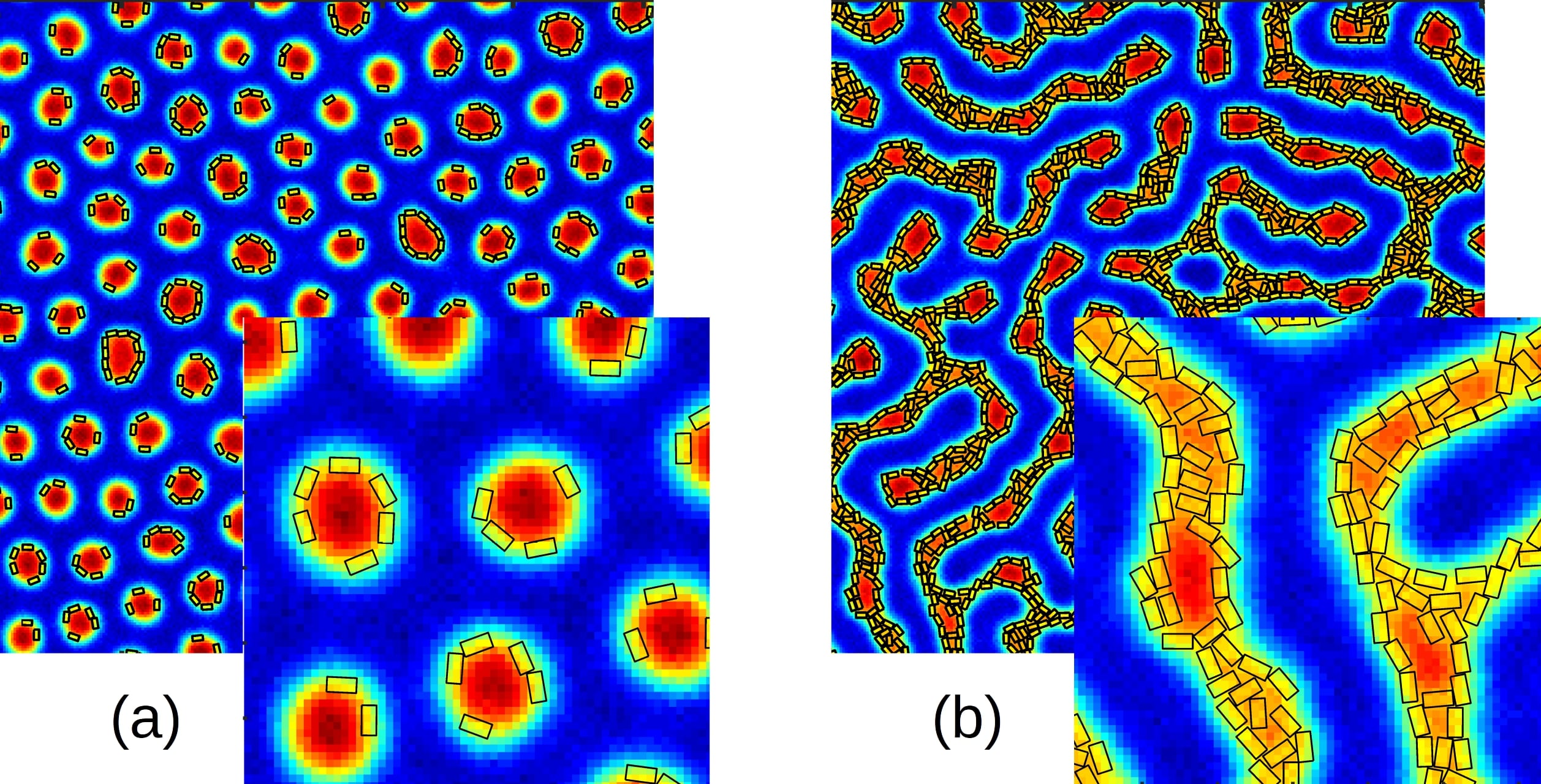}
 \caption{Morphological transition in an asymmetric block copolymer ($f_0=0.35$) induced by the presence of rectangular-shaped particles with size $a_0=2 ,b_0=1 $ (grid point units). The particles have affinity $\psi_0=0$ (neutral) and  the fraction of particles is $\phi_p=0.05$ (a) , $\phi_p=0.2$ (b). }
 \label{fig:anis.rect.trans.asym}
 \end{figure*}

\begin{figure*}[h!]
 \centering
 \includegraphics[width=0.9\textwidth]{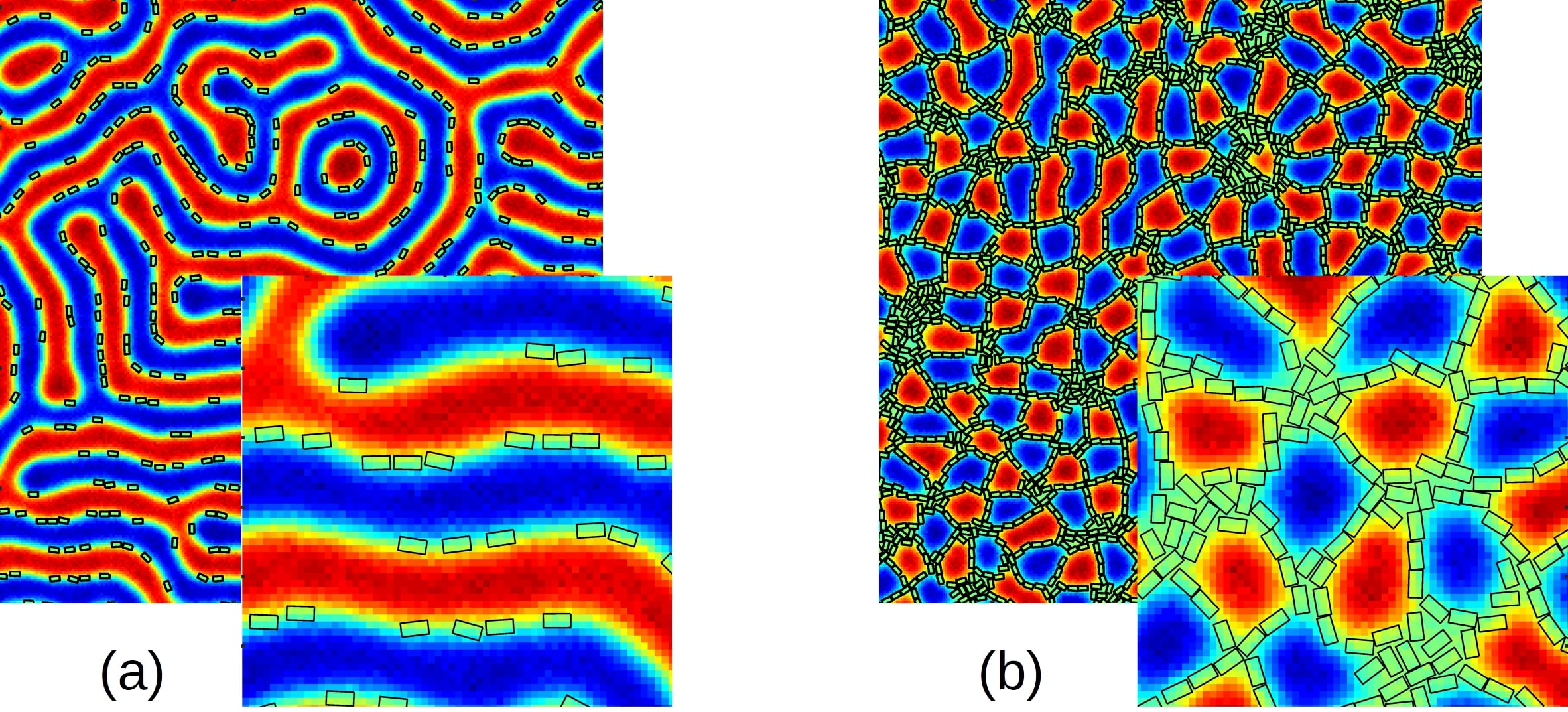}
 \caption{Morphological transition in a symmetric block copolymer ($f_0=0.5$) induced by the presence of rectangular-shaped particles with size $a_0=2 ,b_0=1 $ (grid point units). 
 The particles have affinity $\psi_0=0$ (neutral) and the fraction of particles is $\phi_p=0.05$ (a) , $\phi_p=0.25$ (b). }
 \label{fig:anis.rect.trans.sym}
 \end{figure*}

Homopolymer chains can be grafted to the surface of NPs to create surfactant NPs that tend to segregate at the interface between BCP domains\cite{kim_creating_2007}.
Setting a neutral affinity $\psi_0=0$, we can study the role of anisotropy in the assembly of rectangular NPs ($n=2$). 
Because of the colloidal shape, rectangular nanoparticles tend to orient along the interface, as can be seen in Figures \ref{fig:anis.rect.trans.asym} (a)  for an asymmetric, cylinder-forming BCP. 
In Figure \ref{fig:anis.rect.trans.sym} (a) we can see the orientation of colloids along the lamellar domains. 
In both cases, a large number of particles  oriented in the interface leads to a transition induced by the saturation of the interface and by the formation of bridges along domains. 

Figure \ref{fig:anis.rect.trans.asym} shows the transition from a cylinder-forming BCP with circular domains organised hexagonally (a) into a blue-domain lamellar with a combination of minority (red) phase with colloidal-rich region (b). 
This transition was already reported in simulations of circular NPs and BCP\cite{diaz_phase_2018},   but it can be noticed that many colloids still prefer to assemble at the interface and oriented along it. 

A lamellar, symmetric BCP mixed with rectangular particles can be seen in Figure \ref{fig:anis.rect.trans.sym}. 
In (a), a small concentration of particles is oriented along the interface. 
At larger concentrations the colloids form bridges along domains which break them into smaller, irregular domains. 
The chessboard-like arrangement of BCP domains can be noticed, which was not as clear in mixtures of circles/BCP \cite{diaz_phase_2018}. 
This is clearly a consequence of the anisotropy of colloids, as can be noticed in the detail image in Figure \ref{fig:anis.rect.trans.sym} (b), where the rectangular particles tend to orient along interfaces which can enhance the formation of grid-like configuration of BCP. 
  
\begin{figure*}[h!]
\centering
\includegraphics[width=0.9\textwidth]{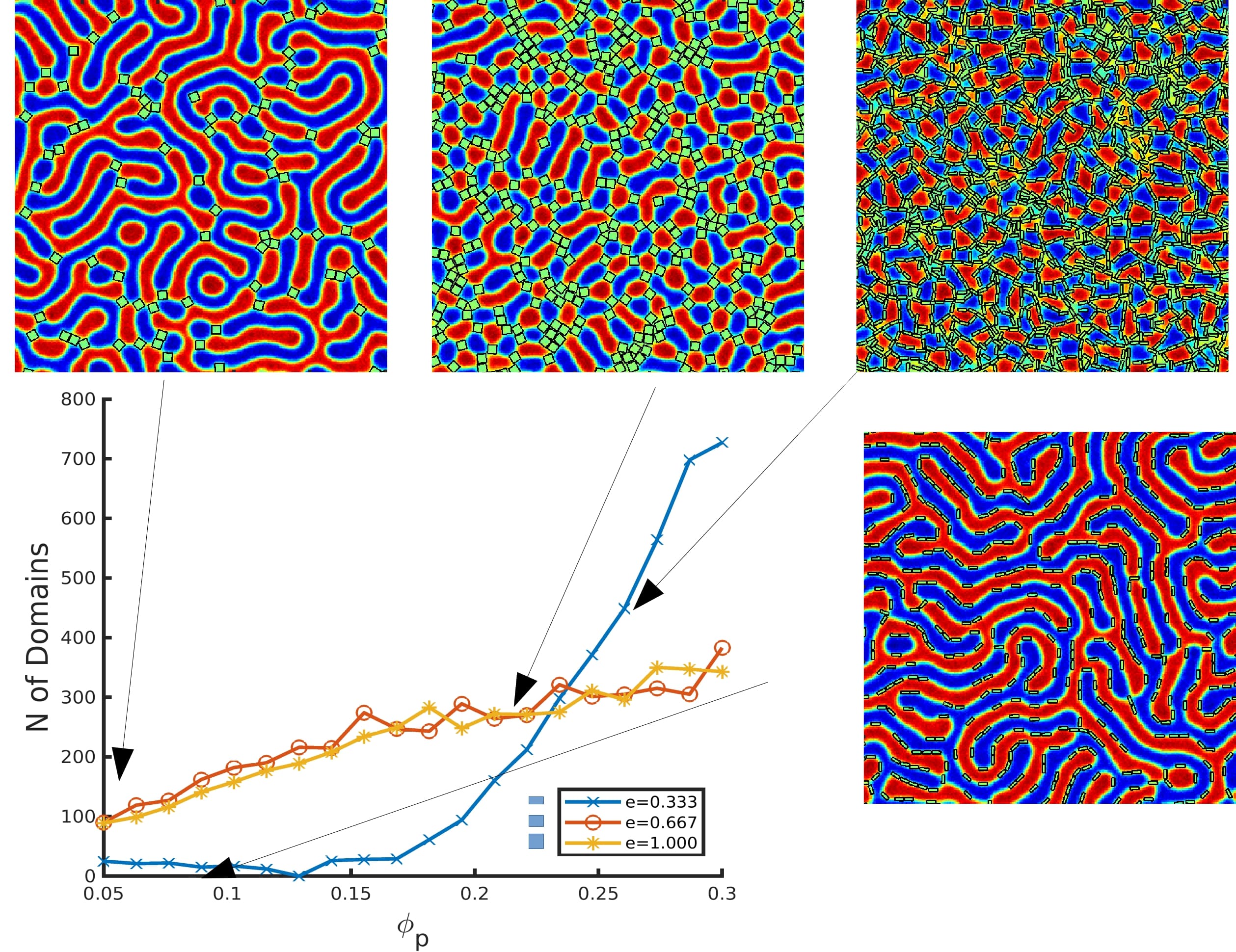}
\caption{Number of BCP domains in terms of the fraction of particles in the system for several particle types. $b=1,2,3$ represent 3 types of rectangular particles with $a=3$ larger semiside and $b$ being the shorter semiside. }
\label{fig:anis.Ndomains}
\end{figure*}

The anisotropy of neutral nanoparticles has been shown to result in alignment along the interface. 
The effect that the aspect ratio of rectangles have on the assembly of colloids and its effect on the overall co-assembly of the BCP composite system can be studied. 
In particular, we can consider a lamellar-forming BCP ($f_0=0.5$) mixed with a concentration $\phi_p$ of neutral nanoparticles with a larger semiaxis sized $a=3$ and an aspect ratio $e=b/a$. 
At low concentration, we can expect NPs to be preferentially segregated towards the interface.
Figure \ref{fig:anis.Ndomains} shows that even at low concentration  weakly anisotropic NPs ($e=0.67$ and $e=1.0$) display a radically distinct behaviour from strongly anisotropic NPs ($e=0.33$). 
The latter are mostly attached to the interface, leaving the BCP largely unmodified. 
This can be tracked by calculating the number of BCP domains in the system, which remains constant up to a high concentration value in which the interface is saturated and the lamellar domains are divided into shorter domains. 
Contrary to that, weakly anisotropic NPs are prone to form aggregates due to their comparatively lower trapping energy at the interface. 
For this reason we BCP morphology suffers a steady increase in the number of domains due to the break-up of domains following the formation of NP clusters.

\subsection{Comparison with experiments}

\begin{figure}[hbtp]
 \centering
 \includegraphics[width=0.6\linewidth]{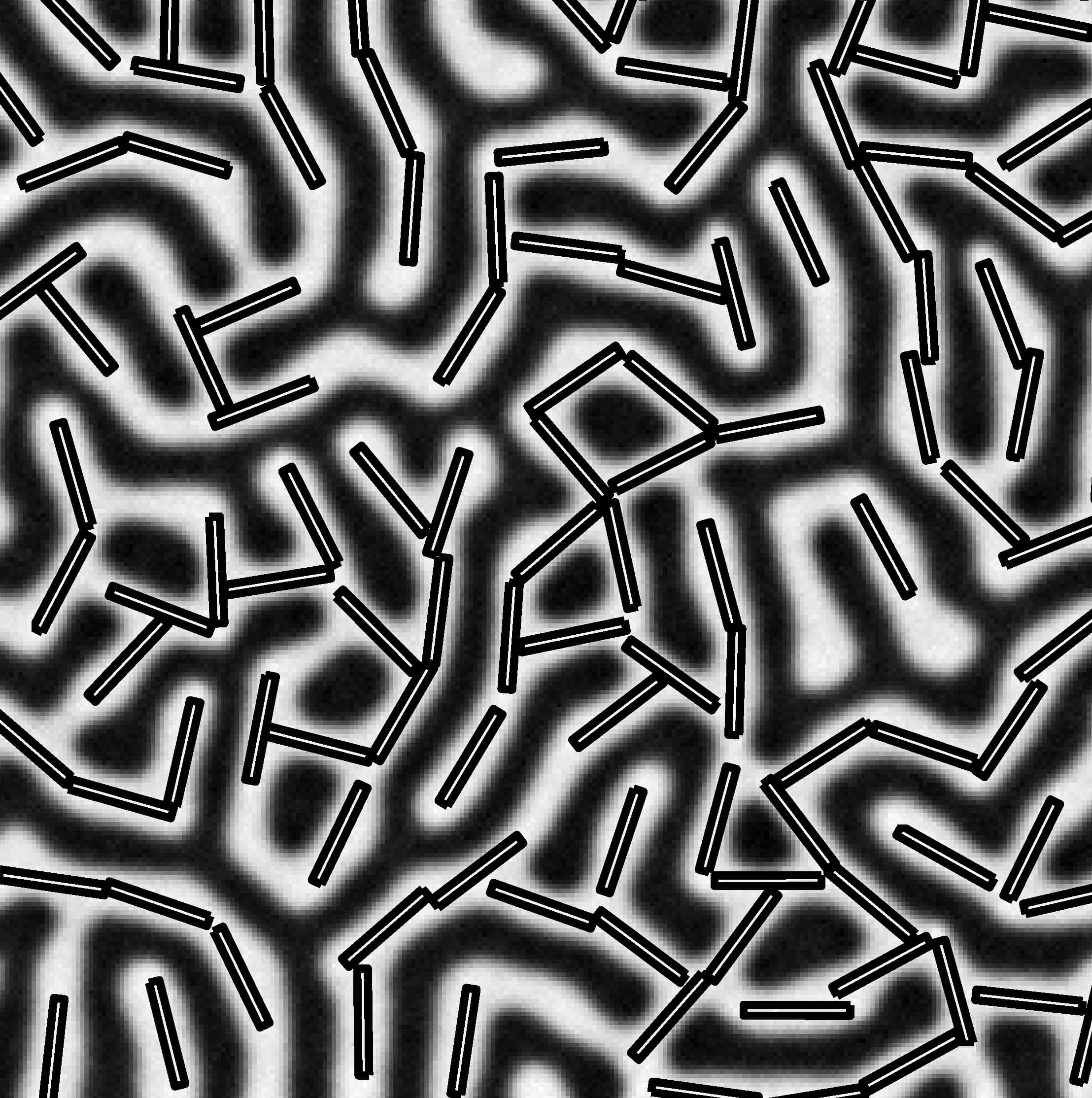}
  \caption{Alignment of rectangular nanoparticles in a symmetric block copolymer.
	  Nanoparticle size is $a_0=12.11$ and $b_0=1.21$ for the major and minor semiaxis while the BCP lamellar spacing is $L\approx 8$. 
   }
  \label{fig:anis.rect.bulk}
 \end{figure}

Anisotropic nanoparticles possess an additional orientational degree of freedom which- when mixed with phase-separating block copolymers- can induce an orientational order. 
Composto et al found nanorods oriented along the lamellar domains \cite{deshmukh_two-dimensional_2007} while end-to-end, in-lamella assembly has been found in thin films \cite{thorkelsson_end--end_2013}. 
 More recently, nanoplates have been found to align within a lamellar-forming PS-\textit{b}-PMMA block copolymer\cite{krook_alignment_2018}. 
While this type of assembly is intrinsically three dimensional, we can aim to study the two-dimensional slice of a nanoplate into a rectangle with sizes $d_1$ and $h$, related to the larger diagonal and the thickness of a nanoplate. 
Along with the lamellar periodicity, we can simulate experimentally relevant parameters of such systems, as in Figure  \ref{fig:anis.rect.bulk} where we observe a clear alignment of rectangular nanoparticles along the lamellar domains. 

\begin{figure}[h!]
\centering
\includegraphics[width=0.6\linewidth]{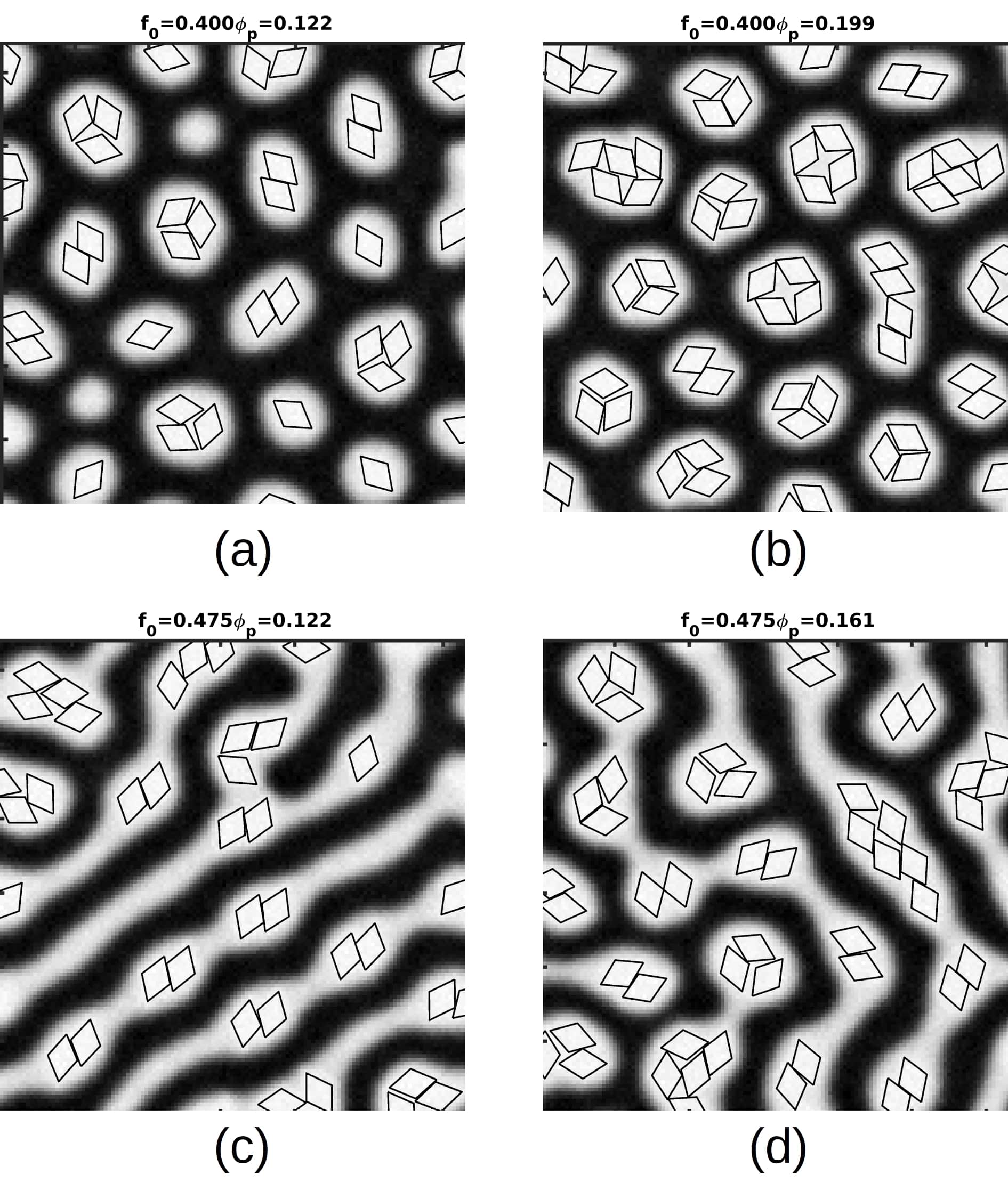}
\caption{Examples of a concentration $\phi_p$ of minority-compatible rhomboid particles mixed with block copolymer with $f_0$ (see titles for values).  }
\label{fig:anis.diamonds.examples}
\end{figure}

The horizontal cross-section of a nanoplate can be simulated as a rhomboid particle using experimental relative sizes \cite{krook_alignment_2018} to simulate realistic rhomboids (nanoplates in 3D) with sizes $b/a=0.629$ (ratio of diagonals) and the relative size between the lamella thickness and larger diagonal tuned accordingly.   
In Figure \ref{fig:anis.diamonds.examples} top-left we can see how the rhomboids are placed within the circular domains, enlarging those in which more than one nanoparticle is placed. Furthermore, the rhomboids tend to assemble in a close-packed manner. 
If we increase the number of particles, this behaviour is even more clear, as in top-right. 
For a lamellar-forming BCP (bottom row), the nanoparticles are not totally confined , only within the stripe domains. 
Because the lamellar thickness cannot accommodate a single NP, they tend to aggregate in order to minimise the distortion induced by the presence of NPs. 

\section{Conclusions}

A general model to simulate anisotropic nanoparticles immersed in block copolymer has been introduced. 
This flexible scheme can cover the family of curves known as super ellipses which include ellipsoids, rectangles and rhomboids. 
A colloid-colloid potential that is proportional to the overlapping area between two objects has been used. 
This analytical potential allows to calculate forces and torques and thus study the time evolution of anisotropic NPs. 
Due to the computational efficiency of the CDS/BD model, this hybrid method can be used to study the co-assembly of BCP/anisotropic nanoparticles, exploring a large parameter space. 

The assembly of square-shaped in a minority phase of a BCP has been studied for different values of concentration and NP size, finding that the enthalpic interaction of the BCP coupling dominates over the assembly of colloids at large particle sizes, compared with the BCP period. 
At high concentrations colloids were able to organise into a side-to-side configuration thanks to the orientational-dependent NP-NP interaction. 
Similarly, minority-compatible rhomboid-shaped NPs where found to align along the domain axis when mixed with a lamellar-forming BCP.

We have been able to compare  experimental conditions by considering both rectangular NPs and rhomboid-shaped nanoparticles in order to mimic three-dimensional nanoplates. 
Using relative sizes from recent experiments \cite{krook_alignment_2018}, we have found alingment of anisotropic colloids within the lamella domains, reproducing  experiments involving nanorods \cite{thorkelsson_end--end_2013,deshmukh_two-dimensional_2007}.

Contrary to A-selective NPs in a A/B BCP, we have studied the co-assembly of neutral, interface-compatible nanoparticles. 
We have found that neutral anisotropic NPs are preferentially aligned along the interface, which results in a saturation of the interface at high concentration. 
This behaviour is nonetheless dependent on the aspect ratio, as a consequence of the strong trapping that anisotropic particles undergo at the interface. 
Weakly anisotropic NPs, on the contrary, have shown a tendency to aggregate which results in a distortion of the lamellar morphology. 

In conclusion, we have presented a general approach to simulate multiple nanoparticle shapes mixed with a BCP matrix. 
Contrary to previous models, non-spherical nanoparticles interact anisotropically both with the surrounding medium and with themselves. 
In combination with an efficient computational method, this allows to simulate high concentration regimes in which the NP-NP interaction is crucial at determining the overall NP configuration.

\section{Acknowledgements}
The work has been performed under the Project HPC-EUROPA3 (INFRAIA-2016-1-730897), with the support of the EC Research Innovation Action under the H2020 Programme; in particular, the authors gratefully acknowledges the computer resources and technical support provided by Barcelona Supercomputing Center (BSC).
I. P. acknowledges support from MINECO (Grant No. PGC2018-098373-B-100), DURSI (Grant No. 2017 SGR 884) and SNF Project No. 200021-175719.

\section{Supporting information}

Supporting information can be found in the file \textit{supp-info.pdf}. Description of the colloid-colloid anisotropic potential.

\bibliography{references} 

\end{document}